\documentclass[prb,twocolumn,showpacs,superscriptaddress]{revtex4}
\usepackage{graphicx}
\usepackage{amsfonts,amsmath,amssymb,float}
\usepackage{bm,dsfont}
\usepackage{color}
\usepackage{bbm}
\usepackage{longtable}
\usepackage[dvips]{epsfig}
\usepackage{amsmath,amssymb,lscape,float}
\usepackage{hyperref}
\usepackage{listings}

\begin{document}
\title{Quantum dynamics of the small-polaron formation 
in a superconducting analog simulator}

\author{Vladimir M. Stojanovi\'c}
\affiliation{Department of Physics, Harvard University,
17 Oxford Street, Cambridge, Massachusetts 02138, USA} 
\affiliation{Department of Physics, University of Belgrade, 
Studentski Trg 12, 11158 Belgrade, Serbia}
\author{Igor Salom}
\affiliation{Institute of Physics Belgrade, University of Belgrade, 
Pregrevica 118, Zemun, Serbia}

\date{\today}

\begin{abstract}
We propose a scheme for investigating the nonequilibrium aspects of small-polaron physics using an array of superconducting 
qubits and microwave resonators. This system, which can be realized with transmon- or gatemon qubits, serves as an analog simulator 
for a lattice model describing a nonlocal coupling of a quantum particle (excitation) to dispersionless phonons. We study its dynamics 
following an excitation-phonon (qubit-resonator) interaction quench using a numerically-exact approach based on a Chebyshev-moment 
expansion of the time-evolution operator of the system. We thereby glean heretofore unavailable insights into the process of small-polaron 
formation resulting from strongly momentum-dependent excitation-phonon interactions, most prominently about its inherent dynamical timescale. 
To further characterize this complex process, we evaluate the excitation-phonon entanglement entropy and show that initially prepared 
bare-excitation Bloch states here dynamically evolve into small-polaron states that are close to being maximally entangled. Finally, by 
computing the dynamical variances of the phonon position- and momentum quadratures we demonstrate a pronounced non-Gaussian character 
of the latter states, with a strong anti-squeezing in both quadratures.
\end{abstract}

\pacs{85.25.Cp, 03.67.Ac, 71.38.Ht}
\maketitle
\section{Introduction}
Recent progress in superconducting (SC) circuits~\cite{VoolDevoretReview:17,SCreviews:17} has enabled 
significant strides in the realm of analog quantum simulation~\cite{Georgescu+:14}. An overwhelming majority of 
proposals for simulating various physical systems using SC circuits is based on arrays of transmon 
qubits and microwave resonators in state-of-the-art circuit quantum electrodynamics (circuit-QED) 
setups~\cite{Koch++:07,GirvinCQEDintro}. Examples include simulators of quantum spin- and spin-boson type systems, 
interacting fermions/bosons, topological states of matter, to name but a few~\cite{Houck+:12,Paraoanu:14,Lamata+:18}. 
In particular, SC simulators of {\em small-polaron} (SP) models~\cite{Mei+:13,Stojanovic+:14,Mostame+:17} 
have proven superior to their counterparts based on trapped ions~\cite{Stojanovic+:12,Mezzacapo+:12}, 
cold polar molecules~\cite{Herrera+Krems:11,Herrera+:13}, and Rydberg atoms/ions~\cite{Hague+MacCormick}. 
Yet, the existing theoretical proposals for simulating SP physics -- not only those based on SC circuits -- solely 
address its static aspects. 

The SP concept captures the physical situation often found in narrow-band semiconductors and insulators, where the motion 
of an itinerant excitation -- an excess charge carrier (electron, hole) or an exciton -- may get hindered by a potential 
well resulting from the host-crystal lattice displacements~\cite{Ranninger:06}. The ensuing SP formation~\cite{AlexandrovDevreese}, 
accompanied by the phonon ``dressing'' of the excitation and an increase in its effective band mass, 
represents the most striking consequence of strong, short-ranged excitation-phonon (e-ph) coupling~\cite{Engelsberg+Schrieffer:63}. 
Yet, some important issues -- e.g, how long it takes for a SP quasiparticle to form following an e-ph interaction 
quench (i.e., a sudden switching-on of the e-ph interaction in a previously uncoupled system) -- remain ill-understood 
as of this writing~\cite{Ku+Trugman:07,Fehske+:11}. On the theoretical side, this fundamental issue remains unresolved 
even in the simplest case of purely local e-ph coupling captured by the time-honored Holstein 
model~\cite{Holstein:59,Wellein+Fehske:97,Wellein+Fehske:98}. On the experimental side, studies of the dynamics of 
polaron formation became possible with advances in ultrafast time-resolved spectroscopies, typically yielding formation 
times of less than a picosecond~\cite{UltrashortExp}.

The compelling need to understand the microscopic mechanisms of charge-carrier transport in complex 
electronic materials -- such as crystalline organic semiconductors~\cite{PeierlsOrganics,Hannewald+:04}, 
semiconducting counterparts of graphene~\cite{PeierlsGraphene}, or cuprates~\cite{Roesch+Gunnarsson:04,Slezak++:06} -- prompted 
investigations of models with strongly momentum-dependent (nonlocal) e-ph interactions~\cite{Stojanovic+:04}. 
Such interactions, whose corresponding vertex functions have explicit dependence on both the excitation 
and phonon quasimomenta, are exemplified by the Peierls-type coupling (also known as Su-Schrieffer-Heeger 
or off-diagonal coupling~\cite{Zoli:04}) that accounts for the dependence of effective excitation hopping amplitudes 
upon phonon states~\cite{Barisic+:70,Stojanovic+Vanevic:08}. Aside from their significance for describing transport 
properties of materials, such couplings have fundamental importance. Namely, they do not obey the Gerlach-L\"{o}wen 
theorem, a formal result that rules out the existence of non-analytical features in the ground-state-related single-particle 
properties for certain classes of coupled e-ph models~\cite{Gerlach+Lowen:91}.

In this paper, motivated by the aforementioned dearth of studies pertaining to the dynamics of SP 
formation, we explore this complex phenomenon using an analog simulator that consists of SC qubits and microwave 
resonators. Adjacent qubits are coupled in this system through a coupler circuit that contains three 
Josephson junctions (JJs). This system, based on transmon qubits, was proposed in the past by one 
of us and collaborators for the purpose of simulating static properties of SP's that originate from nonlocal 
e-ph interactions~\cite{Stojanovic+:14}. Apart from transmons, this system can also be realized with
semiconductor-nanowire-based gatemon qubits~\cite{Larsen++:15,Casparis++:16,Kringhoj+:18}. 

We analyze the time-evolution of SP states ensuing from initially prepared bare-excitation Bloch states with different 
quasimomenta. We do so by combining exact numerical diagonalization of the effective e-ph Hamiltonian of  
the system and the Chebyshev-propagator method~\cite{TalEzer+Kosloff:84,Kosloff:94} for computing its dynamics. 
We determine how the SP formation time after an e-ph interaction quench depends on the initial bare-excitation 
quasimomentum and the e-ph coupling strength. We then further characterize SP formation by evaluating the e-ph
entanglement entropy and showing that at the onset of SP regime it reaches values close to those of maximally-entangled
states. Besides, we evaluate the dynamical variances of the phonon position- and momentum quadratures in SP states 
and demonstrate their pronounced non-Gaussian character, with a substantial anti-squeezing in both quadratures. 
Our findings can be verified in the proposed simulator -- once realized -- by measuring the photon number and 
the attendant squeezing in the resonators.  

The remainder of this paper is organized as follows. In Sec.~\ref{simulator} we present
the layout of our analog simulator and its underlying Hamiltonian. Sec.~\ref{Hamparam}
is set aside for the effective e-ph Hamiltonian of the system, followed by a discussion of 
typical parameter regimes and the salient features of the SP ground state of the system. 
In Sec.~\ref{SysDynMethod} we provide a brief outline of the strategy that we employ to study 
the system dynamics and introduce some relevant timescales in the problem at hand. In Sec.~\ref{ResDiss} 
we present the obtained results for the SP formation time, as well as those found for 
the e-ph entanglement entropy and the dynamical variances of the phonon position-and momentum 
quadratures. We summarize the paper and conclude with some general remarks in Sec.~\ref{SumConcl}. 
Some involved mathematical derivations -- as well as a description of basic aspects of the numerical 
method used and our own implementation thereof -- are relegated to the Appendices~\ref{ExactDiag} -- \ref{CPM}.
\section{System and its Hamiltonian} \label{simulator}
\subsection{Layout of the analog simulator}
\begin{figure}[t!]
\includegraphics[clip,width=8.3cm]{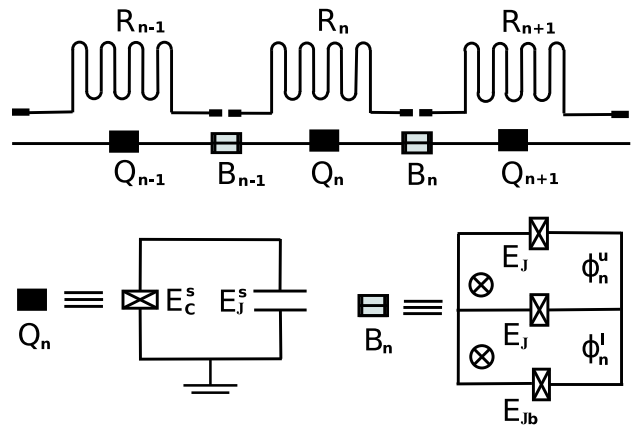}
\caption{\label{fig:circuit}Schematic diagram of the analog-simulator circuit containing SC 
qubits $Q_{n}$ (with charging- and Josephson energies $E^{s}_{C}$ and $E^{s}_{J}$, respectively), 
resonators $R_{n}$, and coupler circuits $B_{n}$ with three Josephson junctions ($n=1,\ldots,N$). 
$\phi_{n}^{l}$ and $\phi_{n}^{u}$ are total fluxes threading the lower and upper loops of $B_{n}$, 
respectively. Qubit $Q_{n}$ interacts with its neighbors through circuits $B_{n-1}$ and $B_n$.}
\end{figure}

The proposed simulator, depicted in Fig.~\ref{fig:circuit}, consists of SC 
qubits ($Q_{n}$) with the energy splitting $\varepsilon_{z}$, microwave resonators 
($R_{n}$) with the photon frequency $\omega_{c}$, and coupler circuits ($B_{n}$) with 
three JJs ($n=1,\ldots,N$). Through the Jordan-Wigner mapping~\cite{ColemanBOOK} the 
pseudospin-$1/2$ degree of freedom of qubits -- represented by the operators 
$\bm{\sigma}_n$ -- plays the role of a spinless-fermion excitation; photons in 
the resonators, created by the operators $a_{n}^{\dagger}$, mimic dispersionless 
phonons. The $n$-th repeating unit of this system is described by the Hamiltonian 
$H_n=H_{n}^{0}+H_{n}^{J}$. Its noninteracting part $H_{n}^{0}$ reads
\begin{equation}
H_{n}^{0}=\frac{\varepsilon_{z}}{2}\:\sigma_{n}^{z}+
\hbar\omega_{c}\:a_{n}^{\dagger}a_{n} \:.
\end{equation}
The Josephson energy of the coupler circuit $B_{n}$~\cite{Geller+:15} -- a generalization of 
a SQUID loop -- is given by
\begin{equation}
H_{n}^{J}=-\sum_{i=1}^{3}E^{i}_{J}\cos\varphi_{n}^{i} \:,
\end{equation}
where $\varphi_{n}^{i}$ are the respective phase drops on the three JJs and $E^{i}_{J}$ 
their corresponding Josephson energies; we henceforth assume that 
$E^{1}_{J}=E^{2}_{J}\equiv E_{J}$ and $E^{3}_{J}=E_{Jb}\neq E_{J}$. 

The qubit- and resonator degrees of freedom are coupled through the flux of the resonator modes that pierces
the upper loops of coupler circuits. The Josephson-coupling energy of the latter circuits, as demonstrated in what 
follows, can be expressed as an $XY$-type (flip-flop) coupling between adjacent qubits with the coupling strength that dynamically 
depends on the resonator (i.e., photon) degrees of freedom. As a result, this indirect inductive-coupling mechanism effectively
gives rise to a qubit-resonator interaction. Besides, coupler circuits are also driven by a microwave radiation (ac 
flux) and subject to an external dc flux. The required ac fluxes can be generated by microwave-pumped control wires 
situated in the vicinity of the respective loops, while the dc flux can be supplied through currents in appropriately 
placed separate control wires.

Let $\phi_{n}^{u}$ and $\phi_{n}^{l}$ be the respective total magnetic fluxes in the upper and lower 
loops of $B_{n}$ (cf. Fig.~\ref{fig:circuit}), both expressed in units of $\Phi_{0}/2\pi$, where 
$\Phi_{0}\equiv hc/(2e)$ is the flux quantum. The upper-loop flux $\phi_{n}^{u}$ includes the ac-driving 
contribution $\pi\cos(\omega_{0}t)$ and one that stems from the resonator modes, i.e.,
\begin{equation}
\phi_{n}^{u}=\pi\cos(\omega_{0}t)+\phi_{n,\textrm{res}} \:,
\end{equation}
where $\phi_{n,\textrm{res}}$ is given by 
\begin{equation}
\phi_{n,\textrm{res}}=\delta\theta[(a_{n+1}+a_{n+1}^{\dagger})-(a_{n}+a_{n}^{\dagger})] \:.
\end{equation}
Here $\delta\theta=[2eA_{\textrm{eff}}/(\hbar d_{0}c)]\times
(\hbar\omega_{c}/C_{0})^{1/2}$, where $A_{\textrm{eff}}$ is the effective coupling area, 
$C_{0}$ the capacitance of the resonator, and $d_{0}$ the effective spacing in the 
resonator~\cite{OrlandoBook}. The lower-loop flux $\phi_{n}^{l}$ also comprises an ac contribution
given by $-(\pi/2)\:\cos(\omega_{0}t)$, with the same frequency as the ac part of $\phi_{n}^{u}$ but
a different amplitude. In addition, it includes a dc part $\phi_{\textrm{dc}}$ -- apart from 
$\omega_{0}$ the only tunable parameter in the system and its main experimental knob. Thus,
$\phi_{n}^{l}$ is given by
\begin{equation}
\phi_{n}^{l}=-\frac{\pi}{2}\:\cos(\omega_{0}t)+\phi_{\textrm{dc}}\:.
\end{equation}
It should be stressed that the amplitudes of the two ac-driving terms are chosen in such a way as 
to ensure that the phase drops $\varphi_n^3$ on the bottom JJs do not have an explicit time 
dependence~\cite{Stojanovic+:14}.

The time dependence of the ac-driving terms makes it natural to carry out further analysis in 
the rotating frame of the drive. While this change of frames leads to a shift in the phonon 
frequency ($\omega_c\rightarrow \delta\omega\equiv\omega_c-\omega_0$), it also renders the 
Josephson-coupling term time dependent. Yet, it can easily be shown that this time dependence 
can be neglected due to its rapidly-oscillating character. The remaining part of that term reads
\begin{equation}\label{eq:HJn_final}
\bar{H}_{n}^{J}=-2\left[t_{r}-\frac{1}{2}E_{J}J_{1}(\pi/2)
\phi_{n,\textrm{res}}\right]\cos(\varphi_{n}-\varphi_{n+1})\:.
\end{equation}
Here $\varphi_{n}$ is the gauge-invariant phase variable of the SC island of the $n$-th qubit~\cite{VoolDevoretReview:17}, 
$J_n(x)$ are Bessel functions of the first kind, and $t_{r}=(E_{Jb}/2)\left(1+\cos\phi_{\textrm{dc}}\right)$, 
where $E_{Jb}$ is chosen to be given by $2E_{J}J_{0}(\pi/2)$. 

In the regime of relevance for transmon/gatemon qubits ($E^{s}_{J}\gg E^{s}_{C}$, where $E^{s}_{C}$ and $E^{s}_{J}$ are 
the charging- and Josephson energies of a single qubit, respectively) $\cos(\varphi_{n}-\varphi_{n+1})$ can be expanded 
up to the second order in $\varphi_{n}-\varphi_{n+1}$. By switching to the pseudospin operators $\bm{\sigma}_n$, it can be 
recast (up to an immaterial additive constant) as $\delta\varphi_{0}^{2}\left[\sigma_{n}^{+}\sigma_{n+1}^{-}+\sigma_{n}^{-}
\sigma_{n+1}^{+}-(\sigma_{n}^{z}+\sigma_{n+1}^{z})/2\right]$. Here $\delta\varphi_{0}^{2}\equiv (2E^{s}_{C}/E^{s}_{J})^{1/2}$, 
hence $\delta\varphi_{0}^{2}\sim 0.15$ for a typical transmon ($E^{s}_{J}/E^{s}_{C}\sim 100$) and 
$\delta\varphi_{0}^{2}\sim 0.28$ for a typical gatemon ($E^{s}_{J}/E^{s}_{C}\sim 25$).

While the original proposal for simulating SP physics with strongly momentum-dependent e-ph interactions 
(Ref.~\onlinecite{Stojanovic+:14}) envisioned the use of transmons -- the most widely used type of SC qubits, 
with superior coherence properties -- it is worthwhile to stress that the system under consideration can also 
be realized with gatemon qubits~\cite{Larsen++:15,Casparis++:16,Kringhoj+:18}. The gatemon is a superconductor-normal-metal-superconductor 
(SNS) type device where an electrostatic gate depletes carriers in a semiconducting weak-link region. This allows 
one to tune the energy of its JJ, and in turn control the qubit frequency~\cite{Larsen++:15}. Because it does
not require an external-flux control, this gate-voltage-controlled counterpart of the transmon has a reduced 
dissipation by a resistive control line and is particularly suitable for use in an external magnetic field. 

Both types of SC qubits under consideration have some advantages with regard to their use in the proposed analog 
simulator. On the one hand, the fact that gatemons do not require external-flux control makes them the preffered 
choice for our present purposes, where the use of external magnetic fluxes is essential (recall Sec.~\ref{simulator} 
above). On the other hand, some other aspects -- e.g., their larger anharmonicity [see Sec.~\ref{characterSP} below] 
and slightly better coherence properties (for a comparison of coherence properties of various SC-qubit types, see 
Ref.~\onlinecite{SCreviews:17}) -- favor the use of transmons. For the sake of completeness, it is worthwhile to add that 
analog simulators of nonlocal e-ph couplings based on other types of SC qubits -- e.g., flux qubits that have large 
anharmonicities -- can also be envisaged, as previously proposed for the local-coupling Holstein model~\cite{Mostame+:17}.
\subsection{Effective coupled excitation-phonon Hamiltonian} \label{Hamparam}
It is pertinent to switch at this point to the spinless-fermion representation of the qubit (pseudospin-$1/2$)
degrees of freedom. The underlying Jordan-Wigner transformation implies that~\cite{ColemanBOOK}
\begin{eqnarray}
\sigma_{n}^{z}&\rightarrow& 2c_{n}^{\dagger}c_{n}-1\:, \nonumber\\
\sigma_{n}^{+}\sigma_{n+1}^{-}+\sigma_{n}^{-}\sigma_{n+1}^{+} 
&\rightarrow& c_{n}^{\dagger}c_{n+1}+\text{h.c.} \:.
\end{eqnarray}
As a consequence, the noninteracting (free) part $H_{\textrm{f}}$ of the effective system 
Hamiltonian -- to be denoted as $H_{\textrm{eff}}=H_{\textrm{f}}+H_{\mathrm{e-ph}}$ in the 
following -- includes the excitation-hopping- and free-phonon terms
\begin{equation}\label{nonint}
H_{\textrm{f}}=-t_{0}(\phi_{\textrm{dc}})\sum^{N}_{n=1}(c_{n}^{\dagger}c_{n+1}
+\textrm{h.c.})+\hbar\delta\omega\sum^{N}_{n=1}a_{n}^{\dagger}a_{n}\:,
\end{equation}
where $t_{0}(\phi_{\textrm{dc}})\equiv 2\delta\varphi_{0}^{2}\:t_r(\phi_{\textrm{dc}})$ is the 
$\phi_{\textrm{dc}}$-dependent bare-excitation hopping integral. [Note that the $\sigma_{n}^{z}$ 
terms from $H_{n}^{0}$ and $\bar{H}_{n}^{J}$ are omitted as they correspond to a constant 
band-energy offset for spinless fermions.] Similarly, the interacting part of $H_{\textrm{eff}}$ 
captures two different mechanisms of nonlocal e-ph interaction and is given by
\begin{eqnarray}\label{Heph} 
H_{\mathrm{e-ph}} &=& g\hbar\delta\omega\:l_0^{-1}\sum^{N}_{n=1}\Big[(c_{n}^{\dagger}
c_{n+1}+\textrm{h.c.})\:\left(u_{n+1}-u_{n}\right) \nonumber \\
&-& c_{n}^{\dagger}c_{n}\left(u_{n+1}-u_{n-1}\right)\Big]\:,  
\end{eqnarray}%
where $g$ is the dimensionless coupling strength and $u_n \equiv l_0(a_{n}+a_{n}^{\dagger})$ the 
local Einstein-phonon displacement at site $n$, with $l_0$ being the phonon zero-point length. The 
first term of $H_{\mathrm{e-ph}}$ corresponds to the Peierls e-ph coupling mechanism, which captures the 
lowest-order (linear) dependence of the excitation hopping amplitude between sites $n$ and $n+1$ on the 
difference $u_{n+1}-u_n$ of the respective phonon displacements~\cite{Stojanovic+Vanevic:08,Hohenadler:16}. 
The other one is the breathing-mode term~\cite{Slezak++:06}, a density-displacement-type mechanism 
which accounts for the antisymmetric coupling of the excitation density $c_{n}^{\dagger}c_{n}$ at site $n$ 
with the phonon displacements on the adjacent sites $n\pm 1$. In other words, it captures a nonlocal phonon-induced 
modulation of the excitation's on-site energy (by contrast to Holstein coupling which describes the 
local phonon-induced modulation of the same energy).

By transforming the e-ph coupling Hamiltonian to its generic momentum-space form
\begin{equation}\label{mscoupling}
H_{\mathrm{e-ph}}=N^{-1/2}\sum_{k,q}\gamma_{\textrm{e-ph}}(k,q)\:
c_{k+q}^{\dagger}c_{k}(a_{-q}^{\dagger}+a_{q}) \:,
\end{equation}
it is straightforward to verify that its corresponding vertex function is given by
\begin{equation}\label{vertex_func}
\gamma_{\textrm{e-ph}}(k,q)=2ig\hbar\delta\omega\:[\:\sin k+\sin q-\sin(k+q)] \:,
\end{equation}
where quasimomenta are expressed in units of the inverse lattice period.
Because this vertex function depends on both the excitation ($k$) and phonon 
($q$) quasimomenta the Hamiltonian $H_{\textrm{eff}}$ does not belong to the realm 
of validity of the Gerlach-L\"{o}wen theorem~\cite{Gerlach+Lowen:91}. As demonstrated 
in Ref.~\onlinecite{Stojanovic+:14} its ground state displays a level-crossing-type 
sharp transition at a critical value of the effective coupling strength 
$\lambda_{\textrm{eff}}\equiv 2g^{2}\:\hbar\delta\omega/t_{0}$. 
While for $\lambda_{\textrm{eff}}<\lambda^{\textrm{c}}_{\textrm{eff}}$ the ground state 
is the (non-degenerate) $K=0$ eigenvalue of the total quasimomentum operator 
\begin{equation}\label{K_tot}
K_{\mathrm{tot}}=\sum_{k}k\:c_{k}^{\dagger}c_{k}+\sum_{q}q\:a_{q}^{\dagger}a_{q} \:,
\end{equation}
for $\lambda_{\textrm{eff}}\ge\lambda^{\textrm{c}}_{\textrm{eff}}$ it is twofold-degenerate
and corresponds to $K=\pm K_{\textrm{gs}}$ (where $K_{\textrm{gs}}\neq 0$ and saturates at 
$\pi/2$ for sufficiently large $\lambda_{\textrm{eff}}$). In this regime the single-particle 
dispersion corresponding to the SP Bloch band has mutually symmetric minima at two nonzero 
quasimomenta, which are here incommensurate with the period of the underlying lattice, a rare 
occurrence in other physical systems~\cite{Stojanovic+:08}.

Aside from a non-analyticity in the ground-state energy of the system, the aforementioned sharp 
transition is manifested by analogous features in the ground-state quasiparticle residue (spectral weight) 
$Z_{\textrm{gs}}\equiv Z_{k=K_{\textrm{gs}}}$, where $Z_{k}\equiv|\langle\Psi_{k}|\psi_{k}\rangle|^{2}$ 
is the module squared of the overlap between the bare-excitation Bloch state 
\begin{equation}\label{bareBloch}
|\Psi_{k}\rangle = c^{\dagger}_{k}|0\rangle_{\textrm{e}}\otimes|0\rangle_{\textrm{ph}}
\end{equation}
and the (dressed) Bloch state $|\psi_{k}\rangle$ of the coupled e-ph system corresponding to the 
same quasimomentum ($K=k$). [Note that $|0\rangle_{\textrm{e}}$ and $|0\rangle_{\textrm{ph}}$ on 
the right-hand-side (RHS) of the last equation stand for the excitation- and phonon vacuum states, 
respectively.] Another quantity characterizing the SP ground state, which shows a non-analiticity 
at $\lambda_{\textrm{eff}}=\lambda^{\textrm{c}}_{\textrm{eff}}$, is the phonon-number expectation 
value in the ground state $\vert{\psi}_{\textrm{gs}}\rangle\equiv\vert{\psi}_{K=K_{\textrm{gs}}}\rangle$:
\begin{equation} \label{phavegs}
\bar{N}_{\text{ph}}={\langle{\psi}_{\textrm{gs}}\vert}\:\sum_{n=1}^{N}
a^{\dagger}_{n}a_{n}\:{\vert{\psi}_{\textrm{gs}}\rangle}\:. 
\end{equation}

The system at hand has another peculiar property -- namely, it is straightforward to demonstrate that the $k=0$ 
bare-excitation Bloch state $|\Psi_{k=0}\rangle$ [cf. Eq.~\eqref{bareBloch}] is an eigenstate of $H_{\textrm{eff}}$ 
for an arbitrary $\lambda_{\textrm{eff}}$, a direct consequence of the fact that the e-ph vertex function 
[cf. Eq.~\eqref{vertex_func}] has the property that $\gamma_{\textrm{e-ph}}(k=0,q)=0$ for any $q$. In particular, 
for $\lambda_{\textrm{eff}}<\lambda^{\textrm{c}}_{\textrm{eff}}$ this state represents the ground state 
of $H_{\textrm{eff}}$.

The relation between the dimensionless coupling strength $g$ and the system-specific 
parameters reads
\begin{equation}\label{gdomega}
g\hbar\delta\omega=\delta\varphi_{0}^{2}\:E_{J}J_1(\pi/2)\delta\theta\:.
\end{equation}
It is worthwhile to notice that $g$ does not depend on the tunable system parameters 
($\omega_0,\:\phi_{\textrm{dc}}$) and we specify it by fixing the product of $\delta\varphi_{0}^{2}$ 
and $E_{J}$ on the RHS of the last equation: $\delta\varphi_{0}^{2}\:E_{J}/2\pi\hbar=100$ GHz.
Given that the typical magnitude of $\delta\varphi_{0}^{2}$ is twice as large for gatemons compared 
to transmons, in a transmon-based realization of this system $E_{J}$ should be taken twice as large 
to retain the same coupling strength and make further discussion completely general. 
Unlike $g$, $\lambda_{\textrm{eff}}$ inherits its dependence on $\phi_{\textrm{dc}}$ 
from $t_{0}$ and is therefore externally tunable:
\begin{equation}\label{expr_lambda}
\lambda_{\textrm{eff}}(\phi_{\textrm{dc}})=g\:\frac{J_1(\pi/2)\delta\theta}
{J_0(\pi/2)\left(1+\cos\phi_{\textrm{dc}}\right)} \:.
\end{equation}
For a typical resonator $\delta\theta\sim 3.5\times 10^{-3}$. Likewise, for $\delta\omega$ we take 
$\delta\omega/2\pi=200 - 300$ MHz. Consequently, for $\delta\omega/2\pi=200$\:MHz 
($300$\:MHz) we obtain $\lambda^{\textrm{c}}_{\textrm{eff}}\approx 0.83\:(0.72)$. 
\section{Dynamics of the small-polaron formation} \label{SysDynMethod}
\subsection{Interaction quench and initial-state preparation}
We study the system dynamics after an e-ph (qubit-resonator) interaction quench at $t=0$, 
assuming that the system was initially prepared in the bare-excitation Bloch state $|\Psi_{k=k_0}\rangle$
with quasimomentum $k_0$. Given that an abrupt change from a bare excitation to a heavily-dressed one here 
takes place for $\lambda_{\textrm{eff}}=\lambda^{\textrm{c}}_{\textrm{eff}}$, a variation of $\phi_{\textrm{dc}}$ 
from slightly below its critical value to slightly above it is equivalent to an interaction quench in this system.

The initial bare-excitation states can be prepared using a general protocol based on an external 
driving and the Rabi coupling between the vacuum state and the desired Bloch state~\cite{Mei+:13}. 
The corresponding preparation time is given by $\tau_{\textrm{prep}}=\pi\hbar/(2\beta_{p})$,
where $\beta_{p}$ is the microwave-pumping amplitude. [Note that an analogous result holds 
in the case of preparing dressed Bloch states -- e.g., a SP ground state -- except that in that 
case the last expression for $\tau_{\textrm{prep}}$ requires another multiplicative factor of $Z^{-1}_{\textrm{gs}}$.]
For a typical pumping amplitude $\beta_{p}/(2\pi\hbar)=10$\:MHz, we obtain $\tau_{\textrm{prep}}= 25$\:ns, 
which is a three orders of magnitude shorter time than currently achievable decoherence times $T_2$ 
of the relevant classes of SC qubits~\cite{SCreviews:17}.
\subsection{Relevant quantities and timescales}
In accordance with the discrete translational symmetry of the system under consideration, its effective Hamiltonian 
commutes with the total quasimomentum operator [cf. Eq.~\eqref{K_tot}], i.e., $[H_{\textrm{eff}},K_{\mathrm{tot}}]=0$.
Therefore, the system evolves within the eigensubspace of $H_{\textrm{eff}}$ 
that corresponds to the eigenvalue $K=k_0$ of $K_{\mathrm{tot}}$. We compute its state $|\psi(t)\rangle$ 
at time $t$ for a simulator with $N=9$ qubits by combining Lanczos-type exact diagonalization~\cite{CullumWilloughbyBook}
of $H_{\textrm{eff}}$ in a symmetry-adapted basis of the truncated Hilbert space of the system (for details, 
see Appendix~\ref{ExactDiag}) and the Chebyshev-propagator method~\cite{TalEzer+Kosloff:84,Kosloff:94}. 
The latter relies on expansions of time-evolution operators into finite series of Chebyshev polynomials of 
the first kind (for general details of this approach and our numerical implementation thereof, see Appendix~\ref{CPM}). 

The knowledge of the state $|\psi(t)\rangle$ of the system at time $t$ allows us to evaluate quantities 
characterizing the ensuing polaronic character of the dressed excitation. One such quantity is the probability 
for the system to remain in the initial state $|\Psi_{k=k_0}\rangle$ at time $t$, given by
\begin{equation} \label{probremain}
P_{k_0}(t)=|\langle\psi(t)|\Psi_{k=k_0}\rangle|^{2}\:.
\end{equation}
This quantity, more precisely the matrix element $\langle\psi(t)|\Psi_{k=k_0}\rangle$, is closely related (up to a 
Fourier transform to the frequency domain) to the momentum-frequency resolved spectral function, a dynamical 
response function that can be extracted in systems of the present type using a generalization of the Ramsey 
interference protocol~\cite{Stojanovic+:14}. Another relevant quantity is the expected total phonon number: 
\begin{equation} \label{phavet}
n_{\textrm{ph}}(t)= \langle\psi(t)|\:\sum_{n=1}^{N}
a^{\dagger}_{n}a_{n}\:|\psi(t)\rangle\:.
\end{equation}
This observable provides a direct quantitative characterization of the dynamical dressing of an
excitation by virtual phonons. In our analog simulator, $n_{\textrm{ph}}(t)$ is amenable to measurement 
by extracting the photon number on the resonators (for details, see Sec.~\ref{characterSP} below).

It is judicious to express the evolution time in units of a timescale closely related to the bare-excitation hopping 
amplitude $t_{0}(\phi_{\textrm{dc}})$. Because the latter here depends on the experimental knob $\phi_{\textrm{dc}}$ 
by design, we choose this timescale to be set by the critical value $\phi^{c}_{\textrm{dc}}=0.972\:\pi$ of $\phi_{\textrm{dc}}$ 
for $\delta\omega/2\pi=300$\:MHz. Thus, the chosen characteristic timescale is 
$\tau_{e,c}\equiv \hbar/t_0(\phi^{c}_{\textrm{dc}})\approx 0.44\:$ns.

One of the most important characteristics of the SP formation process -- yet often elusive in solid-state 
systems exhibiting polaronic behavior~\cite{UltrashortExp} -- is its associated dynamical timescale $\tau_{\textrm{sp}}$. 
It is pertinent to define it as the time at which the phonon dressing (i.e., the phonon-number
expectation value) of an initially bare excitation becomes equal to that of the corresponding 
SP ground state. In other words, $\tau_{\textrm{sp}}$ is defined by the condition 
\begin{equation}\label{SPformcond}
n_{\textrm{ph}}(t=\tau_{\textrm{sp}})= \bar{N}_{\text{ph}}\:,
\end{equation}
where $\bar{N}_{\text{ph}}$ was defined in Eq.~\eqref{phavegs} above. This
ground-state phonon number is in the range $3.9 - 5.1$ ($1.8 - 2$) for 
$\delta\omega/2\pi=200$\:MHz ($300$\:MHz). 
\section{Results and Discussion} \label{ResDiss}
\subsection{Time dependence of $n_{\textrm{ph}}$ and $P_{k_0=\pi/2}$}\label{typquant}
\begin{figure}[t!]
\includegraphics[clip,width=8.3cm]{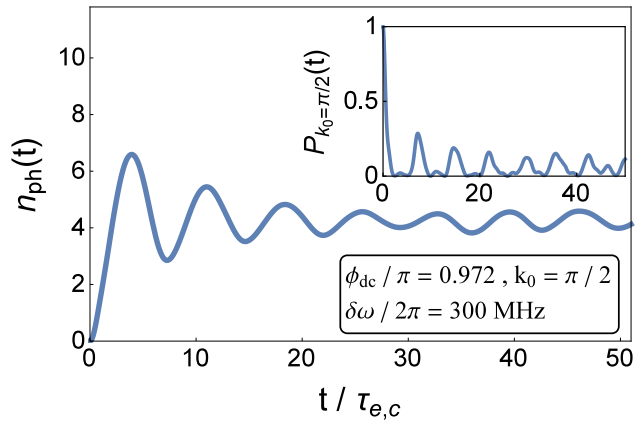}
\caption{\label{fig:nphoft}(Color online) Expected phonon number after an e-ph interaction 
quench at $t=0$ for $k_0=\pi/2$. Inset: the probability to remain in the initial bare-excitation 
Bloch state.}
\end{figure}

Typical results of our numerical calculations of $n_{\textrm{ph}}(t)$ and $P_{k_0=\pi/2}(t)$ are 
presented in Fig.~\ref{fig:nphoft}. They reflect the fact that the system was initially prepared 
in the $|\Psi_{k_0=\pi/2}\rangle$ state, which is not an eigenstate of the system Hamiltonian 
$H_{\textrm{eff}}$ after the quench. 
In fact, for the concrete choice of parameter values used, this state is a superposition of a 
multitude of eigenstates of this Hamiltonian, among which the SP ground state with $K_{\textrm{gs}}=\pi/2$ 
has a weight of only around $0.16$. This explains the presence of dynamical recurrences at later times, i.e., 
a complex oscillatory behavior resulting from the interference of the quantum evolutions of all these eigenstates.

It is instructive to add, for completeness, that our ground-state calculations show 
that for $\lambda_{\textrm{eff}}=\lambda^{\textrm{c}}_{\textrm{eff}}$ -- i.e.,
at the onset of strong-coupling regime in the system under consideration -- there are $3$ discrete states (at each $K$) below 
the one-phonon continuum, while for a larger $\lambda_{\textrm{eff}}$ one can find up to $5$ such states. As a reminder, 
the one-phonon continuum in a coupled e-ph system with gapped (optical-like) phonon modes -- such as, e.g., Einstein-like
phonons in the system at hand -- originates from the onset of the inelastic-scattering threshold at the energy 
$E_{\textrm{gs}}+\hbar\omega_{\textrm{ph}}$ (the minimal energy that a dressed excitation ought to have in order to be able 
to emit a phonon), where $E_{\textrm{gs}}$ is the ground-state energy of the coupled system and $\hbar\omega_{\textrm{ph}}$ 
the energy of one phonon (in our simulator $\omega_{\textrm{ph}}\rightarrow\delta\omega$)~\cite{Engelsberg+Schrieffer:63}. 
The width of this continuum equals the width of the resulting SP Bloch band. Importantly, the discrete (bound) states below the 
one-phonon continuum feature as the coherent part, i.e., sharp peaks in the momentum-frequency resolved spectral function~\cite{Stojanovic+:14}. 
While some details of the dynamics certainly depend on the concrete form of e-ph coupling involved, the increasing number of 
such discrete (split-off from the continuum) states upon increasing coupling strength results in more complex system dynamics.
\subsection{Small-polaron formation time} \label{characterSP}
\begin{figure}[t!]
\includegraphics[clip,width=8.3cm]{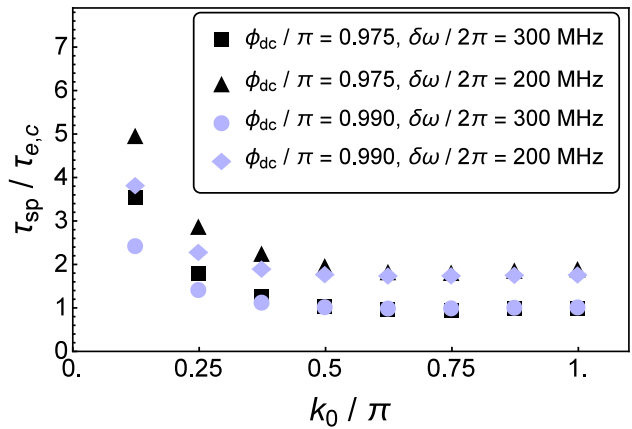}
\caption{\label{fig:formtimeofk}(Color online) SP formation time $\tau_{\textrm{sp}}$ 
for varying initial bare-excitation quasimomenta $k_0$ within the Brillouin zone and 
different choices of values for $\phi_{\textrm{dc}}$ and $\delta\omega$.}
\end{figure}

The dependence of the SP formation time $\tau_{\textrm{sp}}$ on the initial bare-excitation quasimomentum $k_0$ is 
illustrated in Fig.~\ref{fig:formtimeofk} (for symmetry-related reasons, it suffices to consider only quasimomenta 
in one half of the Brillouin zone, i.e., for $0\leq k_0\le\pi$). 
$\tau_{\textrm{sp}}$ clearly shows an upturn for small $k_0$, consistent with the fact that it ought to diverge 
($\tau_{\textrm{sp}}\rightarrow\infty$) as $k_0\rightarrow 0$ because the $k_0=0$ bare-excitation Bloch state is 
an exact eigenstate of $H_{\textrm{eff}}$. Another important feature that can be inferred from the obtained results 
is that $\tau_{\textrm{sp}}$ depends rather weakly on $k_0$ for $\pi/2\leq k_0\le\pi$. This can be contrasted with 
the Holstein-polaron case~\cite{Ku+Trugman:07}, where the analogous dynamical timescale strongly depends on the initial 
bare-excitation quasimomentum.
\begin{figure}[b!]
\includegraphics[clip,width=8.3cm]{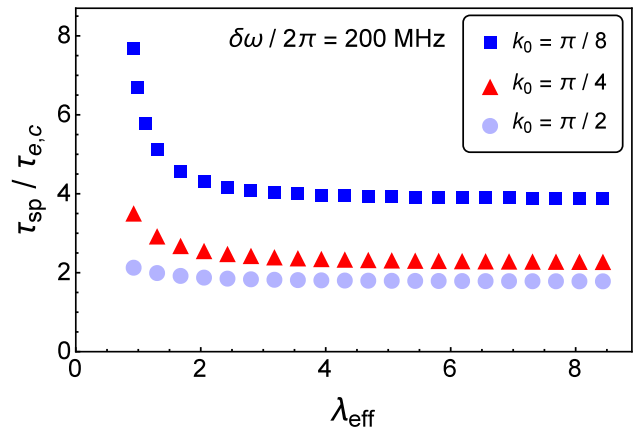}
\caption{\label{fig:formtimeofL}(Color online) Dependence of the SP formation time 
$\tau_{\textrm{sp}}$ on the effective coupling strength $\lambda_{\textrm{eff}}$, shown
for $\delta\omega/2\pi=200$\:MHz and different initial quasimomenta $k_0$.}
\end{figure}

The obtained dependence of $\tau_{\textrm{sp}}$ on the effective coupling strength $\lambda_{\textrm{eff}}$ 
is displayed in Fig.~\ref{fig:formtimeofL}. While it may seem surprising that $\tau_{\textrm{sp}}$ saturates 
for $\lambda_{\textrm{eff}}$ above a threshold value, this actually mimics the behavior of the SP indicators 
(quasiparticle residue, average phonon number) in 
the ground state. In that case a regime of saturation also sets in for $\lambda_{\textrm{eff}}$ slightly above 
its critical value at which a nonanaliticity occurs in all ground-state-related quantities~\cite{Stojanovic+:14}. 
Such a behavior is in stark contrast with that of the momentum-independent Holstein coupling, for which the same 
quantities change monotonously with the coupling strength.

The variation of the SP formation time -- defined by the condition in Eq.~\eqref{SPformcond} -- with the effective coupling 
strength $\lambda_{\textrm{eff}}$ (shown in Fig.~\ref{fig:formtimeofL}) results from two competing tendencies. Namely, with increasing 
$\lambda_{\textrm{eff}}$ phonon dressing of an initially bare excitation becomes faster. However, the average ground-state phonon 
number $\bar{N}_{\text{ph}}$ also becomes larger. In Ref.~\onlinecite{Ku+Trugman:07}, where the dynamics of the Holstein-polaron 
formation were investigated, a regime was observed where the formation time grows with $\lambda_{\textrm{eff}}$ (for weak 
coupling, i.e., small $\lambda_{\textrm{eff}}$) and the one where it decreases (strong coupling, i.e., large 
$\lambda_{\textrm{eff}}$). Despite completely different type of e-ph interaction in the system at hand, we also find such 
regimes. For our typical system parameters, the case with $\delta\omega/2\pi=200$\:MHz [cf. Fig.~\ref{fig:formtimeofL}] corresponds 
to the latter regime, while the case with $\delta\omega/2\pi=300$\:MHz (results not shown here) is characterized by a slow growth 
of $\tau_{\textrm{sp}}$ with increasing $\lambda_{\textrm{eff}}$.

The obtained SP formation times $\tau_{\textrm{sp}}\sim (1-10)\:\tau_{e,c}$ justify {\em a posteriori}
our choice of $\tau_{e,c}$ as the characteristic timescale for the system dynamics. These times are of the order 
of a few nanoseconds and can be verified in this system through photon-number measurements. This is done by 
adding an ancilla qubit (far-detuned from the resonator modes), which couples -- but exclusively during the 
measurement itself -- to a resonator~\cite{Mei+:13}. The photon number on that resonator can then be extracted by
means of a standard quantum non-demolition-measurement readout, which is effectively carried out by measuring 
the transition frequency of the qubit~\cite{Johnson+:10}. The total photon number can then be obtained by 
adding up those found on individual resonators. 

An important issue to address in the context of measuring the photon states in the resonators
is the one pertaining to the anharmonicity $\alpha\equiv E_{12}-E_{01}$ of SC qubits, where 
$E_{ij}$ is the energy difference between states $j$ and $i$ of a single qubit. The anharmonicity
determines the minimal pulse duration $t_p\sim \hbar/|\alpha|$ required to avoid leakage into 
noncomputational single-qubit states. For instance, for a typical transmon with a negative anharmonicity 
of around $200$\:MHz, even microwave pulses with durations on the scale of a few nanoseconds are known to 
be sufficiently frequency selective that one can neglect leakage into higher excited energy levels of the 
transmon and effectively treat it as a two-level system~\cite{GirvinCQEDintro}. For gatemons, whose anharmonicity 
is slightly smaller than that of transmons~\cite{Kringhoj+:18}, similar measurements should also be possible 
for all but the very shortest SP formation times found.
\subsection{Dynamical variances of the phonon position- and momentum quadratures} \label{quadvariance}
It is plausible to expect that nonlocal e-ph correlations in this system are reflected through fluctuations 
within the phonon subsystem, which can be observed via microwave photons in the resonators. To this end, we 
consider the phonon position- and momentum quadratures at an arbitrary -- say $r$-th -- site (in our system 
represented by the photon mode on the $r$-th resonator), defined by the operators $x_r\equiv(a_{r}+a^{\dagger}_{r})/\sqrt{2}$ 
and $p_r\equiv -i(a_{r}-a^{\dagger}_{r})/\sqrt{2}$, respectively. We compute their respective dynamical variances 
$S_x(t)$ and $S_p(t)$, given by
\begin{eqnarray}
S_x(t)&=&\langle\psi(t)|\:x_{r}^2\:|\psi(t)\rangle-\langle\psi(t)|\:x_{r}\:|\psi(t)\rangle^2\:,\nonumber\\
S_p(t)&=&\langle\psi(t)|\:p_{r}^2\:|\psi(t)\rangle-\langle\psi(t)|\:p_{r}\:|\psi(t)\rangle^2\:. \label{variances}
\end{eqnarray}
[Note that, owing to the discrete translational symmetry of the system, the latter quantities should not depend on $r$.] 
The explicit expressions for these variances in our chosen symmetry-adapted basis are provided in Sec.~\ref{matrderiv}).

\begin{figure}[t!]
\includegraphics[clip,width=8.3cm]{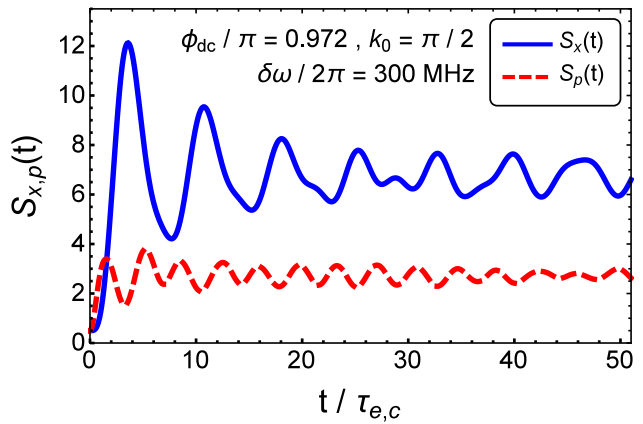}
\caption{\label{fig:dynvar}(Color online) Typical time-dependence of the dynamical variances
$S_x(t)$ and $S_p(t)$. The parameter values are indicated in the plot.}
\end{figure}

Our numerical evaluation of these dynamical variances shows that $S_x$ dominates over $S_p$ at 
all times. For instance, in the weakest-coupling case that yields SP ground state in the system 
at hand -- with $\delta\omega/2\pi=300$\:MHz, and $\phi_{\textrm{dc}}=0.972\:\pi$ (shown in 
Fig.~\ref{fig:dynvar}, which corresponds to $k_0=\pi/2$) -- one finds the maximum of $S_x(t)$ to 
be around $12$. The corresponding anti-squeezing is as large as $13.8$\:dB. 

What can be inferred from Fig.~\ref{fig:dynvar} is that the product $S_x(t)\times S_p(t)$ of the 
two dynamical quadrature variances is consistently much larger than $1/4$, which illustrates a 
pronounced non-Gaussian character of fluctuations within the phonon subsystem. This can be ascribed 
to the nonlocal character of e-ph interaction in this system, with its attendant retardation effects~\cite{Zoli:04}.
\subsection{Dynamics of the excitation-phonon entanglement buildup after the quench}
It is worthwhile to complement our discussion of the dynamics of SP formation 
by evaluating the corresponding e-ph entanglement entropy. This quantity proved
to be very useful in characterizing ground-state properties of SPs -- most prominently
the onset of sharp SP transitions at a critical e-ph coupling strength -- in models with 
strongly momentum-dependent e-ph interactions~\cite{Stojanovic+Vanevic:08}. This motivates 
us to use the same quantity in our present investigation of the SP formation dynamics.

Given that the initial bare-excitation states are of simple-product (separable) character, 
the e-ph entanglement entropy starts its growth from zero at $t=0$.
The density matrix of our composite (bipartite) e-ph system 
at time $t$ is given by
\begin{equation}\label{densmatr}
\rho_{\textrm{e-ph}}(t)=|\psi(t)\rangle\langle\psi(t)| \:.
\end{equation}
The reduced (excitation) density matrix, which has the dimension $N\times N$, 
is obtained by tracing $\rho_{\textrm{e-ph}}(t)$ over the phonon basis:
\begin{equation}\label{reddensmatr}
\rho_{\textrm{e}}(t)=\textrm{Tr}_{\textrm{ph}}
[\rho_{\textrm{e-ph}}(t)]\:.
\end{equation}
The corresponding e-ph entanglement entropy is defined in terms of this last 
reduced density matrix as
\begin{equation}\label{entropyS}
S_{\textrm E}(t)=-\textrm{Tr}_{\textrm{e}}\left[\rho_{\textrm{e}}(t)\ln
\rho_{\textrm{e}}(t)\right]\:.
\end{equation}

The matrix elements of $\rho_{\textrm{e}}(t)$ are computed using Eq.~\eqref{rhoephmatrel}
   [for a detailed derivation of those matrix elements, see Appendix~\ref{entangle}]. 
   The e-ph entanglement entropy in Eq.~\eqref{entropyS} can equivalently be expressed in 
   terms of the eigenvalues $\xi_n(t)$ ($n=1,\ldots,N$) of $\rho_{\textrm{e}}(t)$ (note that 
   $\xi_n>0$ and $\sum^{N}_{n=1}\xi_n=1$):
\begin{equation}
S_{\textrm E}(t)=-\sum^{N}_{n=1}\:\xi_n(t)\:\ln\xi_n(t)\:.
\end{equation}
Generally speaking, the maximal value that can be reached by this quantity is 
\begin{equation}
S_{\textrm{max-ent}}=\ln N\:,
\end{equation}
obtained when $\xi_n=N^{-1}$ for each $n$ (maximally-entangled states).

\begin{figure}[t!]
\includegraphics[clip,width=8.3cm]{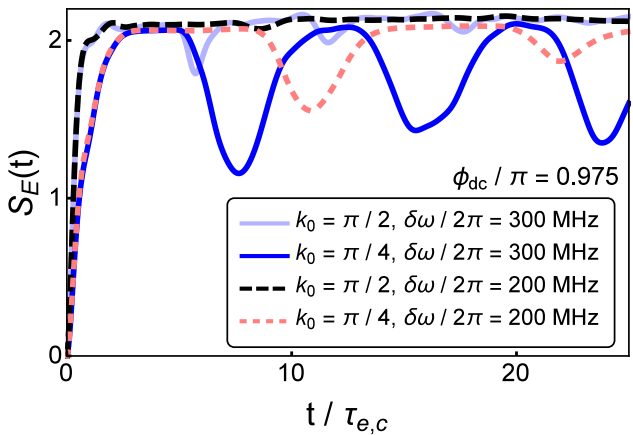}
\caption{\label{fig:dyn_entFull}(Color online) Time-dependence of the e-ph entanglement entropy 
for $\phi_{\textrm{dc}}=0.975\:\pi$ and different choices of values for $k_0$ and $\delta\omega$.}
\end{figure}

\begin{figure}[b!]
\includegraphics[clip,width=8.3cm]{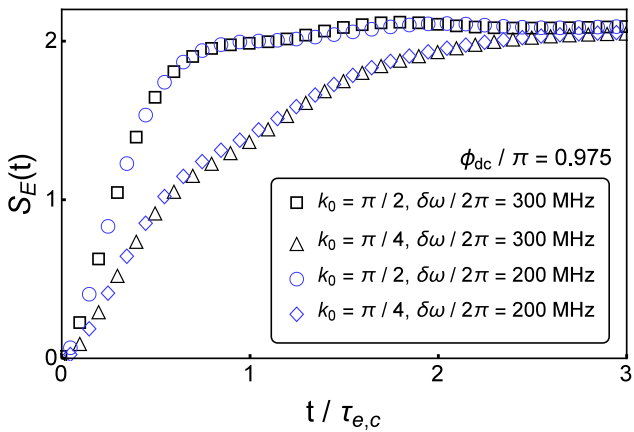}
\caption{\label{fig:dyn_entZoom}(Color online) Short-time behavior of the e-ph entanglement entropy 
for $\phi_{\textrm{dc}}=0.975\:\pi$ and different choices of values for $k_0$ and $\delta\omega$.}
\end{figure}

Our explicit evaluation of the e-ph entanglement entropy is illustrated in Figs.~\ref{fig:dyn_entFull} and
\ref{fig:dyn_entZoom}, where it is depicted for $\phi_{\textrm{dc}}=0.975\:\pi$ and two different initial 
bare-excitation quasimomenta ($k_0=\pi/2,\:\pi/4$) and phonon frequencies ($\delta\omega/2\pi=200$\:MHz, 
\:$300$\:MHz).   
In particular, Fig.~\ref{fig:dyn_entFull} illustrates that the growth of this entanglement 
entropy from zero at $t=0$ starts with an abrupt increase on timescales of the order of a few $\tau_{e,c}$.
This short-time behavior of the entropy is depicted separately in Fig.~\ref{fig:dyn_entZoom}, from which we 
can infer that at short times $S_{\textrm E}(t)$ depends 
on $k_0$, but is essentially independent of $\delta\omega$. The abrupt increase of $S_{\textrm E}(t)$ is followed by oscillations 
at later times. Those oscillations, which are much more pronounced for $k_0=\pi/4$ than for $k_0=\pi/2$, are another manifestation 
of the late-time recurrences, akin to those found in $n_{\textrm{ph}}(t)$ and $P_{k_0=\pi/2}(t)$ [recall Sec.~\ref{typquant}].

Another important feature of the e-ph entanglement entropy, which can be inferred from the obtained results, is that at times 
$t\approx (2-3) \:\tau_{e,c}$ -- coinciding with the corresponding SP formation times $\tau_{\textrm{sp}}$ -- this quantity
indeed reaches values close to those characterizing maximally-entangled states [note that for $N=9$, we have $S_{\textrm{max-ent}}=2.197$]. 
For instance, the respective maximal values of $S_{\textrm E}(t)$ obtained for the above case of $k_0=\pi/2$ are $2.141$ for 
$\delta\omega/2\pi=200$\:MHz and $2.115$ for $300$\:MHz. This is consistent with the results of an earlier study that reached 
the conclusion that typical SP ground states are essentially maximally entangled~\cite{Stojanovic+Vanevic:08}. 
\section{Summary and Conclusions} \label{SumConcl}
To summarize, in this work we explored the dynamics of small-polaron formation in the presence of two different mechanisms of nonlocal 
excitation-phonon interaction within the framework of an analog simulator. In this simulator, which is based on an array of coupled 
superconducting qubits (transmons or gatemons) and microwave resonators, the pseudospin degree of freedom of qubits plays the 
role of spinless-fermion excitation, while photons in the resonators mimick dispersionless phonons. By employing a numerically-exact 
approch -- diagonalization of the effective system Hamiltonian combined with the Chebyshev-propagator method for computing its 
dynamics -- we determined the formation time of small polarons that ensue from initially-prepared bare-excitation Bloch states 
following an excitation-phonon (qubit-resonator) interaction quench. 

We analyzed how this important dynamical timescale depends on the initial bare-excitation quasimomentum and the effective coupling 
strength. We then further characterized the system dynamics by evaluating the excitation-phonon entanglement entropy and demonstrating 
its growth from zero (before the interaction quench) to values close to those inherent to maximally-entangled states. Finally, by 
computing the dynamical variances of the phonon position- and momentum quadratures we also demonstrated the non-Gaussian character of 
small-polaron states resulting from the quench, with a strong anti-squeezing in both quadratures.

The present work constitutes a systematic theoretical study of the quantum dynamics of small-polaron 
formation resulting from strongly momentum-dependent excitation-phonon coupling. Such couplings, in their own right, are of utmost 
importance for understanding charge-transport mechanisms in several classes of electronic materials. The advanced measurement capabilities 
of the proposed superconducting analog simulator should allow an accurate verification of our quantitative predictions. To make contact with 
previous studies of the same phenomenon involving other types of polarons, we compared and contrasted our findings with those pertaining to
the small-polaron formation dynamics in the presence of purely local (momentum-independent) Holstein-type excitation-phonon 
interaction~\cite{Ku+Trugman:07,Fehske+:11}. We found, for instance, that in the system at hand -- where excitation-phonon coupling 
itself is strongly momentum-dependent -- the small-polaron formation time shows a weaker dependence on the initial bare-excitation 
quasimomentum than in the Holstein-polaron case. 

Several directions of future work can be envisioned. Firstly, the proposed simulator can be utilized for investigations of further 
nonequilibrium aspects of small-polaron physics, which have so far also been discussed only for Holstein-type excitation-phonon 
interaction~\cite{Vidmar+:11,Golez+:12,Sayyad+Eckstein:15,Dorfner+:15}. Examples of such aspects include the small-polaron dynamics in the 
presence of an external electric field~\cite{Vidmar+:11}, as well as the dynamics following a strong oscillatory pulse~\cite{Golez+:12}. 
Furthermore, while the proposed system serves as a simulator for a one-dimensional excitation-phonon model, the continuously-improving scalability 
of superconducting-qubit systems should allow one to fabricate -- in not-too-distant future -- a two-dimensional counterpart of this simulator. 
Such a system could be used for studying the effects of dimensionality on the formation of small-polaron-type quasiparticles; analogous effects 
have proven to be quite interesting in the case of Holstein polarons. Finally, a different type of qubit-resonator arrays -- featuring effective 
$XXZ$-type coupling between qubits~\cite{Heule+++} -- would allow an investigation of intersite bipolarons~\cite{RanningerBipolaron:06}, quasiparticles 
closely related to small polarons. Experimental realization of the proposed system is keenly anticipated.
\begin{acknowledgments}
V.M.S. acknowledges useful discussions with L. Tian during previous collaborations 
on related topics. In the initial stages of this project V.M.S. was supported by the 
SNSF. This work was supported in part by the Serbian Ministry of Science and 
Technological Development under grant numbers 171027 (V.M.S.) and ON 171031 (I.S.).
\end{acknowledgments}
\appendix 
\section{Symmetry-adapted basis and details of exact diagonalization} \label{ExactDiag}
The Hilbert space of the coupled e-ph system is spanned by states $|n\rangle_e\otimes|\mathbf{m}\rangle_\text{ph}$,
where $|n\rangle_e\equiv c_{n}^{\dagger}|0\rangle_e$ corresponds to the excitation 
localized at the site $n$, while 
\begin{equation}\label{mphvect}
|\mathbf{m}\rangle_\text{ph} = \prod_{n=1}^{N\otimes}
\frac{(b_n^\dagger)^{m_n}}{\sqrt{m_n!}}\:|0\rangle_\text{ph}\:,
\end{equation}
where $\mathbf{m}\equiv (m_1,\ldots,m_N)$ are the phonon occupation numbers at different sites. 
We restrict ourselves to the truncated phonon Hilbert space that includes states with the total 
number of phonons $m=\sum_{n=1}^N m_n$ not larger than $M$, where $0\le m_n \le m$. Accordingly, 
the total Hilbert space of the system has the dimension $D = D_\text{e} \times D_\text{ph}$, where 
$D_\text{e} = N$ and $D_\text{ph}=(M+N)!/(M!N!)$.

The dimension of the Hamiltonian matrix to be diagonalized can be further reduced
by exploiting the discrete translational symmetry of the system, whose 
mathematical expression is the commutation $[H_{\textrm{eff}},K_{\mathrm{tot}}]=0$
of operators $H_{\textrm{eff}}$ and $K_{\mathrm{tot}}$. This permits diagonalization 
of $H_{\textrm{eff}}$ in Hilbert-space sectors corresponding to the eigen-subspaces 
of $K_{\mathrm{tot}}$, where the dimension of each of those $K$-sectors of the total Hilbert 
space coincides with that of the truncated phonon space, i.e., $D_{K}=D_{\textrm{ph}}$. 
To this end, we utilize the symmetry-adapted basis 
\begin{equation}\label{symmbasis}
|K,\mathbf{m}\rangle = N^{-1/2} \sum_{n=1}^N e^{iKn}\,\mathcal{T}_n(|1\rangle_\text{e} 
\otimes |\mathbf{m}\rangle_\text{ph}) \:,
\end{equation}
with $\mathcal{T}_{n}$ being the (discrete) translation operators whose
action ought to comply with the periodic boundary conditions. The last equation
can be recast as
\begin{equation}\label{symmbasalter}
|K,\mathbf{m}\rangle = N^{-1/2} \sum_{n=1}^N e^{iKn}\, |n\rangle_\text{e} 
\otimes \mathcal{T}^{\textrm{ph}}_n|\mathbf{m}\rangle_\text{ph} \:,
\end{equation}
where the operators $\mathcal{T}^{\textrm{ph}}_n$ represent the action of discrete
translations in the phonon Hilbert space. Note that, if $|\mathbf{m}\rangle_\text{ph}$
is defined by a set of occupation numbers 
\begin{equation}\label{mphvectors}
|\mathbf{m}\rangle_\text{ph} = |m_1,m_2,\ldots,m_{N}\rangle_\text{ph} \:,
\end{equation}
then $\mathcal{T}^{\textrm{ph}}_n\:|\mathbf{m}\rangle_\text{ph}\equiv
|\mathcal{T}^{\textrm{ph}}_{n}\mathbf{m}\rangle$ is given by
\begin{equation}\label{mphvecttransl}
|\mathcal{T}^{\textrm{ph}}_{n}\mathbf{m}\rangle = 
|m_{N-n+1},m_{N-n+2},\ldots,m_{N-n}\rangle_\text{ph} \:.
\end{equation}
In general, in terms of the original phonon occupation numbers, the $r$-th occupation number 
in $|\mathcal{T}^{\textrm{ph}}_{n}\mathbf{m}\rangle$ is given by $m_{s(r,n)}$, where the site
index $s(r,n)$ is defined by
\begin{equation}\label{findef}
s(r,n)\equiv\begin{cases} 
         N-n+r\:, \:\text{for $r\le n$} \\
         r-n\:, \: \text{for $r>n$}
        \end{cases}\:.
\end{equation}

Regarding the ground-state calculations, we follow an established phonon Hilbert-space truncation 
procedure~\cite{Wellein+Fehske:97} whereby the system size ($N$) and the maximum number of phonons 
retained ($M$) are increased until the convergence for the ground-state energy and the phonon distribution 
is reached. Our adopted convergence criterion is that the relative error in the ground-state energy and 
the phonon distribution upon further increase of $N$ and $M$ is not larger than $10^{-4}$. The adopted 
criterion is here satisfied for the system size $N=9$ (with periodic boundary conditions) and requires 
the total of $M=10$ phonons. 
\section{Matrix elements and expectation values} \label{matrderiv}
\subsection{Derivation of the matrix elements of relevant observables} 
In what follows, we first derive the expressions for the expectation values of a generic
observable with respect to the state $|\psi(t)\rangle$ of the system at time $t$.
In view of our use of the symmetry-adapted basis [cf. Eq.~\eqref{symmbasis}],  
we do so by deriving the matrix elements of the same observables in that basis. 
We then specialize to the relevant observables for our present work -- the total 
phonon (photon) number [defined by Eq.~\eqref{phavet}], as well as the variances 
of the phonon position and momentum quadratures corresponding to an arbitrary site 
[defined by Eq.~\eqref{variances}].

We start from the decomposition of the state $|\psi(t)\rangle$ in the symmetry-adapted 
basis [defined in Eq.~\eqref{symmbasis} above]
\begin{equation}\label{decomp}
|\psi(t)\rangle=\sum_{\mathbf{m}}C^{K}_{\mathbf{m}}
(t)|K,\mathbf{m}\rangle \:,
\end{equation}
where the expansion coefficients $C^{K}_{\mathbf{m}}(t)$ can be obtained through our 
computation of the state evolution. For an arbitrary observable $A$ we then have
\begin{equation}\label{geneq}
\langle \psi(t)|A|\psi(t)\rangle=\sum_{\mathbf{m},\mathbf{m'}}
C^{K*}_{\mathbf{m'}}(t)C^{K}_{\mathbf{m}}(t)
\langle K,\mathbf{m'}|A|K,\mathbf{m}\rangle \:,
\end{equation}
which -- with already known expansion coefficients -- leaves us with the task of 
calculating the matrix elements $\langle K,\mathbf{m'}|\:A\:|K,\mathbf{m}\rangle$ 
for the relevant observables. 

Assuming -- as is the case for our relevant observables -- that $A$ depends only on 
phonon operators, it is straightforward to show, using Eq.~\eqref{symmbasalter}, that
\begin{equation}\label{matrixel}
\langle K,\mathbf{m'}|\:A\:|K,\mathbf{m}\rangle 
=\frac{1}{N}\sum_{n=1}^{N}\:\langle\mathcal{T}^{\textrm{ph}}_{n}
\mathbf{m'}|\:A\:|\mathcal{T}^{\textrm{ph}}_{n}\mathbf{m}\rangle\:,
\end{equation}
where in deriving this last result we made use of the fact that 
${}_{\textrm{e}}\langle n'|n\rangle_{\textrm{e}}=\delta_{nn'}$. 
Before embarking on further derivations it is useful to note that 
$\langle\mathcal{T}^{\textrm{ph}}_{n}\mathbf{m'}|\mathcal{T}^{\textrm{ph}}_{n}
\mathbf{m}\rangle$ is independent of $n$ and equals $1$ if the two sets of 
phonon occupation numbers -- $\mathbf{m}$ and $\mathbf{m'}$ -- are completely 
the same, otherwise it evaluates to zero. In other words,
\begin{equation}\label{iprodtrans}
\langle\mathcal{T}^{\textrm{ph}}_{n}\mathbf{m'}|\mathcal{T}^{\textrm{ph}}_{n}
\mathbf{m}\rangle=\delta_{\mathbf{m},\mathbf{m'}}\:.
\end{equation}

In the simplest case, for $A=a^{\dagger}_{r}a_{r}$, we first note that
\begin{equation}\label{}
a^{\dagger}_{r}a_{r}\:|\mathcal{T}^{\textrm{ph}}_{n}\mathbf{m}\rangle
=m_{s(r,n)}\:|\mathcal{T}^{\textrm{ph}}_{n}\mathbf{m}\rangle \:,
\end{equation}
where $s(i,n)$ is defined by Eq.~\eqref{findef}. By inserting the last result 
into Eq.~\eqref{matrixel} and making use of Eq.~\eqref{iprodtrans} we then easily 
obtain that 
\begin{equation}\label{}
\langle K,\mathbf{m'}|\:a^{\dagger}_{r}a_{r}\:|K,\mathbf{m}\rangle 
=\frac{\delta_{\mathbf{m},\mathbf{m'}}}{N}\:\sum_{n=1}^{N}\:m_{n}\:.
\end{equation}
where -- owing to the discrete translational symmetry of the system -- the 
RHS of the last equation does not explicitly depend on $r$. [In writing the last 
equation, we made use of the fact that $\sum_{n=1}^{N}\:m_{s(r,n)}\equiv\sum_{n=1}^{N}\:m_{n}$]. 
By extension, for $A=\sum_{r=1}^{N}\:a^{\dagger}_{r}a_{r}$ (total photon number), we get
\begin{equation}\label{}
\langle K,\mathbf{m'}|\sum_{r=1}^{N}\:a^{\dagger}_{r}a_{r}|K,\mathbf{m}
\rangle=\delta_{\mathbf{m},\mathbf{m'}}\:\sum_{n=1}^{N}\:m_{n}\:.
\end{equation}
Upon inserting the last result into the general Eq.~\eqref{geneq}, we obtain 
the desired expectation value
\begin{equation}\label{}
\langle \psi(t)|\sum_{r=1}^{N}\:a^{\dagger}_{r}a_{r}|\psi(t)\rangle=
\:\sum_{n,\mathbf{m}}\:m_n\:|C^{K}_{\mathbf{m}}(t)|^{2} \:.
\end{equation}

For $A=a_{r}$, we first notice that 
\begin{equation}\label{}
a_{r}|\mathcal{T}^{\textrm{ph}}_{n}\mathbf{m}\rangle=
\sqrt{\displaystyle m_{s(r,n)}}\:|\mathcal{T}^{\textrm{ph}}_{n}
\mathbf{m}_{(r,-1)}\rangle \:.
\end{equation}
where $|\mathcal{T}^{\textrm{ph}}_{n}\mathbf{m}_{(r,-1)}\rangle$ is the vector obtained 
by changing the $r$-th occupation number in $|\mathcal{T}^{\textrm{ph}}_{n}\mathbf{m}\rangle$
from $m_{s(r,n)}$ to $m_{s(r,n)}-1$. This implies that
\begin{equation}\label{tnatn}
\langle\mathcal{T}^{\textrm{ph}}_{n}\mathbf{m'}|\:a_{r}\:|\mathcal{T}^{\textrm{ph}}_{n}
\mathbf{m}\rangle =\sqrt{\displaystyle m_{s(r,n)}}
\end{equation}
provided that the two sets ($\mathbf{m}$ and $\mathbf{m'}$) have the same 
occupation numbers except at site $s(r,n)$ where $m'_{s(r,n)}$ should be equal 
to $m_{s(r,n)}-1$; otherwise, $\langle\mathcal{T}^{\textrm{ph}}_{n}\mathbf{m'}|
\:a_{r}\:|\mathcal{T}^{\textrm{ph}}_{n}\mathbf{m}\rangle = 0$.
The desired matrix element $\langle K,\mathbf{m'}|\:a_{r}\:|K,\mathbf{m}\rangle$ is 
obtained by combining Eq.~\eqref{tnatn} and the general result in Eq.~\eqref{matrixel}.

In an analogous fashion, for $A=a^{\dagger}_{r}$ we obtain 
\begin{equation}\label{tnadtn}
\langle\mathcal{T}^{\textrm{ph}}_{n}\mathbf{m'}|\:a^{\dagger}_{r}\:|
\mathcal{T}^{\textrm{ph}}_{n}\mathbf{m}\rangle =\:\sqrt{\displaystyle m_{s(r,n)}+1}
\end{equation}
if the two sets ($\mathbf{m}$ and $\mathbf{m'}$) have the same 
occupation numbers except at site $s(r,n)$ where $m'_{s(r,n)}$ should be equal 
to $m_{s(r,n)}+1$; otherwise, $\langle\mathcal{T}^{\textrm{ph}}_{n}\mathbf{m'}|\:a^{\dagger}_{r}\:|
\mathcal{T}^{\textrm{ph}}_{n}\mathbf{m}\rangle=0$. The matrix element 
$\langle K,\mathbf{m'}|\:a^{\dagger}_{r}\:|K,\mathbf{m}\rangle$ sought for is easily obtained 
by inserting the expression in Eq.~\eqref{tnadtn} into the general Eq.~\eqref{matrixel}.

By combining the derived expressions for $\langle K,\mathbf{m'}|\:a_{r}\:|K,\mathbf{m}\rangle$ 
and $\langle K,\mathbf{m'}|\:a^{\dagger}_{r}\:|K,\mathbf{m}\rangle$, we can easily obtain the 
desired results for $\langle K,\mathbf{m'}|\:x_{r}\:|K,\mathbf{m}\rangle$ and 
$\langle K,\mathbf{m'}|\:p_{r}\:|K,\mathbf{m}\rangle$.

When $A=x_r^{2}=(a^{\dagger}_{r}+a_{r})^2/2$ or $A=p_r^{2}=-(a^{\dagger}_{r}-a_{r})^2/2$, 
we first note that 
\begin{eqnarray}\label{}
x_r^{2}&\equiv&\frac{1}{2}\:[2a^{\dagger}_{r}a_{r}+1+(a^{\dagger}_{r})^2+a_{r}^2] \:,\\
p_r^{2}&\equiv&\frac{1}{2}\:[2a^{\dagger}_{r}a_{r}+1-(a^{\dagger}_{r})^2-a_{r}^2]\:.
\end{eqnarray}
Repeating the above procedure, to compute the desired matrix elements 
$\langle K,\mathbf{m'}|\:(a^{\dagger}_{r})^2\:|K,\mathbf{m}\rangle$ and 
$\langle K,\mathbf{m'}|\:a^2_{r}\:|K,\mathbf{m}\rangle$, we have to first
determine $\langle\mathcal{T}^{\textrm{ph}}_{n}\mathbf{m'}|\:(a^{\dagger}_{r})^2\:|
\mathcal{T}^{\textrm{ph}}_{n}\mathbf{m}\rangle$ and 
$\langle\mathcal{T}^{\textrm{ph}}_{n}\mathbf{m'}|\:a_{r}^2\:|
\mathcal{T}^{\textrm{ph}}_{n}\mathbf{m}\rangle$.
It is straightforward to show that, for instance, 
\begin{equation}\label{adsmel}
\langle\mathcal{T}^{\textrm{ph}}_{n}\mathbf{m'}|\:(a^{\dagger}_{r})^2\:|
\mathcal{T}^{\textrm{ph}}_{n}\mathbf{m}\rangle= \sqrt{[m_{s(r,n)}+1]
[m_{s(r,n)}+2]} 
\end{equation}
provided that the two sets ($\mathbf{m}$ and $\mathbf{m'}$) have the same 
occupation numbers except at site $s(r,n)$ where $m'_{s(r,n)}$ should be equal to $m_{s(r,n)}+2$;
otherwise, $\langle K,\mathbf{m'}|\:(a^{\dagger}_{r})^2\:|K,\mathbf{m}\rangle= 0$.
Similarly, we have that
\begin{equation}\label{asmel}
\langle\mathcal{T}^{\textrm{ph}}_{n}\mathbf{m'}|\:a_{r}^2\:|
\mathcal{T}^{\textrm{ph}}_{n}\mathbf{m}\rangle=\sqrt{m_{s(r,n)}[m_{s(r,n)}-1]} 
\end{equation}
if the two sets ($\mathbf{m}$ and $\mathbf{m'}$) have the same 
occupation numbers except at site $s(r,n)$ where $m'_{s(r,n)}$ should be 
equal to $m_{s(r,n)}-2$; otherwise, $\langle\mathcal{T}^{\textrm{ph}}_{n}\mathbf{m'}|\:a_{r}^2\:|
\mathcal{T}^{\textrm{ph}}_{n}\mathbf{m}\rangle=0$.

With the aid of the expressions for the matrix elements obtained thus far, and using the general expression 
in Eq.~\eqref{geneq} with the coefficients $C^{K}_{\mathbf{m}}(t)$ obtained from the computation of the 
system dynamics, one can straightforwardly obtain the variances $S_x(t)$ and $S_p(t)$ of the position 
and momentum quadratures [cf. Eq.~\eqref{variances}].
\subsection{Derivation of the matrix elements of the reduced density matrix} \label{entangle}
In what follows, we derive expressions for the matrix elements of the reduced
density matrix assuming that the system under consideration evolves starting 
from a bare-excitation Bloch state with quasimomentum $k_0$ at $t=0$.

We make use of our standard symmetry-adapted basis [cf. Eq.~\eqref{symmbasalter}] 
for $K=k_0$:
\begin{equation}\label{symmbas_k0}
|K=k_0,\mathbf{m}\rangle = N^{-1/2} \sum_{n=1}^N e^{ik_0 n}\,|n\rangle_\text{e} 
\otimes \mathcal{T}^{\textrm{ph}}_n|\mathbf{m}\rangle_\text{ph} \:.
\end{equation}
and start by expanding the state $|\psi(t)\rangle$ of the system at time $t$ with 
respect to this basis:
\begin{equation}\label{decomp_k0}
|\psi(t)\rangle=\sum_{\mathbf{m}}C^{k_0}_{\mathbf{m}}(t)|k_0,\mathbf{m}\rangle \:,
\end{equation}
The density matrix of our composite (bipartite) e-ph system at time $t$ is given by
Eq.~\eqref{densmatr} and -- using the expansion in Eq.~\eqref{decomp_k0} -- can be 
expressed as
\begin{equation}
\rho_{\textrm{e-ph}}(t)=\sum_{\mathbf{m},\mathbf{m'}}\:C^{k_0*}_{\mathbf{m'}}(t)
\:C^{k_0}_{\mathbf{m}}(t)\:|k_0,\mathbf{m}\rangle\langle k_0,\mathbf{m'}|\:.
\end{equation}
By now making use of Eq.~\eqref{symmbasalter}, i.e., its special case for 
$K=k_0$, we further obtain:
\begin{eqnarray}\label{rhoeph}
\rho_{\textrm{e-ph}}(t)&=& N^{-1}\sum_{\mathbf{m},\mathbf{m'}}\sum^{N}_{n,n'=1}\:
e^{ik_0(n-n')}C^{k_0 *}_{\mathbf{m'}}(t)\:C^{k_0}_{\mathbf{m}}(t)\: \nonumber\\
&\times&|n\rangle\langle n'|\otimes|\mathcal{T}^{\textrm{ph}}_{n}\mathbf{m}\rangle
\langle \mathcal{T}^{\textrm{ph}}_{n'}\mathbf{m'}|\:.
\end{eqnarray}

The reduced excitation density matrix is obtained by tracing the last density matrix 
over the phonon basis [cf. Eq.~\eqref{reddensmatr}]. Let $\mathbf{m''}$ be the dummy 
index for the phonon basis states, i.e., the set of all phonon occupation-number configurations. 
Then we have
\begin{equation}
\rho_{\textrm{e}}(t)=\sum_{\mathbf{m''}}\langle \mathbf{m''}|\:
\rho_{\textrm{e-ph}}(t)\:|\mathbf{m''}\rangle\:,
\end{equation}
which, by inserting $\rho_{\textrm{e-ph}}(t)$ from Eq.~\eqref{rhoeph}, becomes
\begin{eqnarray}\label{exprrhoeph}
\rho_{\textrm{e}}(t)&=&N^{-1}\sum_{\mathbf{m},\mathbf{m'},\mathbf{m''}}
\sum^{N}_{n,n'=1}\:e^{ik_0(n-n')}C^{k_0 *}_{\mathbf{m'}}(t)\:C^{k_0}_{\mathbf{m}}(t)\:\nonumber\\
&\times& \:\langle\mathcal{T}^{\textrm{ph}}_{n'}\mathbf{m'}|\mathbf{m''}\rangle   
\langle \mathbf{m''}|\mathcal{T}^{\textrm{ph}}_{n}\mathbf{m}\rangle\:
|n\rangle\langle n'|\:.
\end{eqnarray}
We now note that
\begin{equation}\label{interm}
\sum_{\mathbf{m''}}\langle\mathcal{T}^{\textrm{ph}}_{n'}\mathbf{m'}|\mathbf{m''}\rangle   
\langle \mathbf{m''}|\mathcal{T}^{\textrm{ph}}_{n}\mathbf{m}\rangle=
\langle\mathcal{T}^{\textrm{ph}}_{n'}\mathbf{m'}|\mathcal{T}^{\textrm{ph}}_{n}
\mathbf{m}\rangle \:,
\end{equation}
where we made use of the completeness relation in the phonon Hilbert space
\begin{equation}\label{}
\sum_{\mathbf{m''}}|\mathbf{m''}\rangle \langle \mathbf{m''}|=\mathbbm{1}\:.
\end{equation}
Using the result in Eq.~\eqref{interm}, the expression for $\rho_{\textrm{e}}(t)$ 
in Eq.~\eqref{exprrhoeph} now simplifies to
\begin{eqnarray}\label{rhoephfinal}
\rho_{\textrm{e}} (t)&=& N^{-1}\sum_{\mathbf{m},\mathbf{m'}}\sum^{N}_{n,n'=1}\:
e^{ik_0(n-n')}C^{k_0 *}_{\mathbf{m'}}(t)\:C^{k_0}_{\mathbf{m}}(t)\:\nonumber\\
&\times& \:\langle\mathcal{T}^{\textrm{ph}}_{n'}\mathbf{m'}|\mathcal{T}^{\textrm{ph}}_{n}
\mathbf{m}\rangle\:|n\rangle\langle n'|\:.
\end{eqnarray}
From the last equation we readily read off the final expression for the matrix elements 
of the reduced excitation density matrix:
\begin{eqnarray}\label{rhoephmatrel}
\left(\rho_{\textrm{e}}\right)_{nn'}(t)&=&N^{-1}\:e^{ik_0(n-n')}\sum_{\mathbf{m},\mathbf{m'}}
\:C^{k_0 *}_{\mathbf{m'}}(t)\:C^{k_0}_{\mathbf{m}}(t) \nonumber \\
&\times&\:\langle\mathbf{m'}\:|\:\mathcal{T}^{\textrm{ph}}_{n-n'}\mathbf{m}\rangle\:,
\end{eqnarray}
where we made use of the fact that $\langle\mathcal{T}^{\textrm{ph}}_{n'}\mathbf{m'}\:|\:
\mathcal{T}^{\textrm{ph}}_{n}\mathbf{m}\rangle\equiv \langle\mathbf{m'}\:|\:
\mathcal{T}^{\textrm{ph}}_{n-n'}\mathbf{m}\rangle$. It is also useful to note that 
the final result in Eq.~\eqref{rhoephmatrel} can more succinctly be recast as
\begin{equation}\label{rhoephmatrelconc}
\left(\rho_{\textrm{e}}\right)_{nn'}(t)= N^{-1}\:e^{ik_0(n-n')}\:
\langle\psi(t)\:|\mathcal{T}^{\textrm{ph}}_{n-n'}|\:\psi(t)\rangle\:.
\end{equation}

In order to evaluate the matrix element $\langle\mathbf{m'}\:|\:\mathcal{T}^{\textrm{ph}}_{n-n'}\mathbf{m}\rangle$, 
it is useful to recall Eqs.~\eqref{mphvecttransl} and \eqref{findef}. Note that 
$\langle\mathcal{T}^{\textrm{ph}}_{n'}\mathbf{m'}\:|\:\mathcal{T}^{\textrm{ph}}_{n}\mathbf{m}\rangle=1$ 
if all the corresponding phonon occupation numbers in $|\mathcal{T}^{\textrm{ph}}_{n}\mathbf{m}\rangle$
and $|\mathcal{T}^{\textrm{ph}}_{n'}\mathbf{m'}\rangle$ are the same, otherwise this matrix element
evaluates to zero. 
\section{Chebyshev-propagator method (CPM) for dynamics}\label{CPM}
In the following, we briefly recapitulate the essential aspects of the computational
technique utilized in the present work -- the Chebyshev-propagator method 
(CPM)~\cite{TalEzer+Kosloff:84} -- followed by some basic details of our concrete 
implementation thereof. A more detailed introduction into the CPM is provided in Ref.~\onlinecite{Kosloff:94}.
\subsection{Basics of the CPM}\label{CPMbasics}
For a system described by the Hamiltonian $H$, the time-evolution operator 
$U(t+\delta t,t)=U(\delta t)=e^{-iH\delta t}$ can be expanded in a finite 
series of $N_{\textrm{C}}$ Chebyshev polynomials of the first-kind 
$T_p(x)=\cos(p\arccos x)$~\cite{TalEzer+Kosloff:84,Kosloff:94}:
\begin{equation}\label{chebyshev}
U(\delta t)=e^{-ib\delta t}\left[c_{0}(a\delta t)+2\sum_{p=1}^{N_{\textrm{C}}} 
c_p(a\delta t)T_p(\widetilde{H}) \right] \:.
\end{equation}
Here $\widetilde{H}=(H-b)/a$ is a rescaled Hamiltonian of the system, where
$E_{\textrm{min}}$ ($E_{\textrm{max}}$) is the minimal (maximal) eigenvalue 
of $H$, $b=(E_{\textrm{max}}+E_{\textrm{min}})/2$, and 
$a=(E_{\textrm{max}}-E_{\textrm{min}}+\epsilon)/2$, with 
$\epsilon=\alpha_{\textrm{c}}(E_{\textrm{max}}-E_{\textrm{min}})$ being 
introduced to ensure that the rescaled spectrum lies well inside $[-1,1]$.
The expansion coefficients are given by 
\begin{equation}
c_p(a\delta t)=\int^{1}_{-1}\frac{T_p(x)\:e^{-ixt}}
{\pi\sqrt{1-x^2}}\:dx=(-i)^p\:J_p(a\delta t) \:, 
\end{equation}
where $J_p(a\delta t)$ is the $p$-th order Bessel function of the first kind. 
In cases where the system Hamiltonian does not depend explicitly on time, these 
expansion coefficients also depend only on the time step $\delta t$ (but not 
explicitly on time $t$), thus it is sufficient to compute them only once. 

The recurrence relations for the Chebyshev polynomials~\cite{TalEzer+Kosloff:84} 
can be used to simplify the computation of the state evolution 
$|\psi(t+\delta t)\rangle=U(\delta t)|\psi(t)\rangle$ 
from one point on a time grid to the next one. The problem 
is effectively reduced to the iterative evaluation of vectors 
$|v_p\rangle\equiv T_p(\widetilde{H})|\psi(t)\rangle$ using the recursive relation 
\begin{equation}\label{rec_vectors}
|v_{p+1}\rangle=\widetilde{H}|v_p\rangle-|v_{p-1}\rangle \:,
\end{equation}
where $|v_0\rangle=\psi(t)$ and $|v_{1}\rangle=\widetilde{H}|v_0\rangle$.
Evolving the state vector $|\psi(t)\rangle$ from one time step to the next one requires $N_{\textrm{C}}$ 
matrix-vector multiplications of a given complex vector with a sparse Hamiltonian matrix, a step that 
for the system evolution from $t = 0$ to $t=t_f$ has to be performed $t_f/\delta t$ times.

Given that the CPM requires only the knowledge of two extremal eigenvalues of the system
Hamiltonian, it is convenient to combine it with Lanczos-type diagonalization for sparse 
matrices~\cite{CullumWilloughbyBook}. The CPM has by now proven to be superior to other 
direct or iterative integration schemes, in terms of both computational cost and accuracy~\cite{Leforestier+:91}.
\subsection{Implementation details and numerical consistency checks}
The results obtained for the system dynamics in this work were based on calculations performed for a system 
with $N=9$ qubits, with up to $M=20$ phonons in the truncated phonon Hilbert space. Thus, the resulting maximal 
dimension of the truncated phonon Hilbert space (as discussed in Sec.~\ref{ExactDiag} of this Appendix, this is 
also the dimension of any $K$-sector of the full Hilbert space) was $D\approx 10^7$ and to make the storage of the 
nonzero matrix elements possible we used the sparse-matrix form. Reaching the numerical convergence in our dynamics 
calculations typically required us to use between $N_{\textrm{C}}=9$ and $N_{\textrm{C}}=14$ Chebyshev polynomials 
in the expansion given by Eq.~\eqref{chebyshev}. Besides, the smallest time step required for numerical convergence 
was $\delta t = 0.05\:\tau_{\textrm{e,c}}$, i.e., up to $20$ time steps were used within the period that corresponds 
to the physically meaningful (excitation-hopping) timescale $\tau_{\textrm{e,c}}\approx\:0.44\:$ns. Our runs included 
those with the total evolution times $t_f$ as large as $100\:\tau_{\textrm{e,c}}$, i.e., with up to $2,000$ such time 
steps. In our calculations, $\epsilon$ was kept at the fixed value of $10^{-3}$ (cf. Sec.~\ref{CPMbasics}).

We carried out our CPM-based dynamics calculations on an $8$-core, $3.5$\:GHz Intel Xeon CPU E5-1620 machine, with 
a total of $32$\:GB of main memory. The runs that were required to obtain all the results presented in this paper 
consumed less than $250$ CPU hours. 

The results were checked for consistency whenever it was possible. In particular, testing for unitarity turned out to be 
a good measure of convergence of the CPM. Namely, at each iteration step the norm of the evolving state vector was calculated 
and any deviation from unity larger than $10^{-4}$ was considered a surefire sign that a higher precision (i.e., either
a shorter time step or a larger $N_{\textrm{C}}$) is needed. Despite the fact that we maintained this unitarity margin of 
error to be much lower than $10^{-4}$ throughout our calculations, this was not always sufficient and additional convergence 
tests -- performed by increasing computational precision and confirming the stability of the results -- were carried out.

As another internal consistency check, we used the mathematical relation between the expectation values of $x_r^2$, $p_r^2$, 
and the phonon-number operator $a^{\dagger}_{r}a_{r}$ that stems from the identity $x_r^2 + p_r^2 = 2a^{\dagger}_{r}a_{r} + 1$. 
This relation was satisfied by our data at $10^{-7}$ precision, which is a highly nontrivial test as the expectation 
values of $x_r^2$ and $p_r^2$ on one hand, and that of $a^{\dagger}_{r}a_{r}$ on the other, were evaluated by completely 
different and mutually independent means.

\end{document}